\documentclass[twocolumn,showpacs,amsmath,amssymb,aps,prl,nobalancelastpage]{revtex4}

\usepackage{amsthm}
\usepackage{newlfont}
\usepackage{graphicx}
\usepackage{subfigure}
\usepackage[section]{placeins}
\usepackage{psfrag}
\usepackage{caption2}
\usepackage{flafter}
\usepackage{bm}%
\def\fnum@figure{\figurename\thefigure}
\renewcommand{\figurename}{Fig.}

\newcommand{\pa}{\partial}

\begin{document}


\title{State transition induced by  self-steepening and self phase-modulation}
\author{J.S.~He$^{1,2}$, S.W.~Xu$^{3}$, M.S.~Ruderman$^{4}$} \author{R.~Erd\'elyi$^{4}$}
\address{$^{1}$Department of Mathematics, Ningbo University, Ningbo, Zhejiang 315211, P.R.China.\\
$^{2}$ DAMTP, University of Cambridge, Cambridge CB3 0WA, UK\\
$^{3}$School of Mathematics, USTC, Hefei, Anhui 230026, P.R.China\\
$^{4}$Solar Physics and Space Plasma Research Centre, University of
Sheffield, Sheffield, S3 7RH, UK}
\begin{abstract}
We present a rational solution for a mixed nonlinear Schr\"odinger
(MNLS) equation. This solution  has two free parameters $a$ and $b$
representing the contributions of  self-steepening and self
phase-modulation (SPM) of an associated physical system. It
describes five soliton states:  a  paired bright-bright soliton,
single soliton, a paired bright-grey  soliton, a paired bright-black
soliton, and a rogue wave state. We show that the transition among
these five states is induced by self-steepening and SPM through
tuning the values of $a$ and $b$. This is a unique and potentially
fundamentally important phenomenon in a physical system described by
the MNLS equation.\\
 \noindent \textbf{KEYWORDS}: Mixed nonlinear Schr\"odinger equation,
 state transfer, Alfv\'en wave, ultra-short light pulse,
 rational solution.
\end{abstract}
\pacs{ 05.45.Yv, 42.65.Re, 52.35.Sb,02.30.Ik }
\maketitle \mbox{\vspace{-1cm}}
 \noindent{ \bf Introduction.} The mixed nonlinear Schr\"odinger (MNLS)
 equation \cite{KMio},
\begin{equation}\label{MDNLS}
q_{t}-iq_{xx}+a(q^{\ast}q^2)_{x}+ib q^{\ast}q^2=0,
\end{equation}
has been derived by the modified reductive perturbation method to
describe the propagation of  e.g. the Alfv\'en waves with small but
finite amplitude  along the magnetic field in the cold plasma
approximation widely applicable in solar, solar-terrestrial, space
and astrophysics\cite{medvedev,sulem}. Later it has been shown
\cite{Tzoar,Anderson,Zabolotskii} that the MNLS equation also
provides an accurate modelling of ultra-short light pulse
propagation in optical fibers. In eq.~(\ref{MDNLS}) the complex
quantity $q$ represents the magnetic field perturbation in the case
of Alfv\'en waves, and the electrical field envelope in the case of
waves in optical fibres. The asterisk denotes complex conjugate, $a$
and $b$ are two non-negative constants determined by the unperturbed
state, and the subscript $x$ ($t$) denotes  the
 partial derivative with respect to the spatial coordinate $x$ (time $t$).
The MNLS equation reduces to the Nonlinear Schr\"odinger (NLS)
equation when $a=0$, and to the Derivative Nonlinear Schr\"odinger
(DNLS) equation when $b=0$. Because the MNLS equation is applicable
even in the case where the lengths of the envelope wave and the
carrier wave are comparable \cite{KMio}, it is more general than the
nonlinear Schr\"odinger equation in the modelling of waves in
optical fibres. The last three terms in eq.~(\ref{MDNLS}) describe
the group velocity dispersion (GVD), self-steepening and self
phase-modulation (SPM), respectively.

The Lax pair for the MNLS equation provides the mathematical basis
for the solvability of this equation by the inverse scattering
method and Darboux transformation (DT), the latter being defined by
the Wadati-Konno-Ichikawa (WKI) spectral problem and the first
non-trivial flow \cite{MikiWadati1}
$$
\begin{array}{ll}
\pa_{x}\psi=(-aJ\lambda^2+Q_1\lambda+Q_0)\psi \equiv U\psi, \vspace*{2mm}\\
\pa_{t}\psi=(-2a^2J\lambda^4+V_{3}\lambda^3+V_{2}\lambda^2 +
V_{1}\lambda+V_0)\psi \equiv V\psi,
\end{array}
$$
with the reduction condition $r=-q^\ast$\/. Here the complex
quantity $\lambda$ is the eigenvalue (or spectral parameter), and
$\psi=(\phi,\varphi)^T$ is the eigenfunction associated with
$\lambda$\/. The superscript $T$ denotes transposition. There are
also other methods to show the integrability of the MNLS equation
and to obtain its exact solutions \cite{AKundu,
Seenuvasakumaran,QingDing,Kundu}. Various types of solutions of the
MNLS equation, including a soliton and a breather have already been
obtained in \cite{TKawata,Chowdhury,mihalache,Doktorov2,Rangwala,
OCWrigh, TiechengXia,HaiQiangZhang,Shanliang,MinLi}. The decay of
soliton solution for a
 perturbed MNLS system has been demonstrated numerically \cite{serkindecay}. Small
 perturbations of the MNLS equation have been studied either by a direct
  method \cite{perturbation}, or using the inverse scattering
  transform \cite{Shchesnovich,VMLashkin}.

Up to now, all known solutions of the MNLS equation represent rather
common nonlinear wave solutions, like solitons and breathers. These
are similar corresponding ones to their ancestors --- the NLS and
DNLS equation, so they do not describe any specific properties of
either the Alfv\'en waves in magnetised plasmas, or ultrashort light
pulses in optical fibres. Thus, it is a long-standing problem to
find unique phenomena related to the simultaneous effects of
self-steepening  and SPM  that can be described by the explicit
analytical solutions of the MNLS equation.

In this letter we present novel rational solutions of the MNLS
equation. These solutions describe five states of the associated
systems: a paired bright-bright soliton state, a single soliton
state, a paired bright-grey soliton state, a  paired bright-dark
soliton state, and a rogue wave state. We also show that the
transition between these five states is induced by the
self-steepening and SPM through tuning the values of $a$ and $b$\
determined  by the physical properties of the background state.

\noindent{\bf Analytical form of rational solution.} By the one-fold
DT and Taylor-expansion, according to a similar procedure
demonstrated by  \cite{he2,xuhe,xxuhe,taohe,hexup} for constructing
the rational rogue wave (RW) of the NLS, DNLS, Hirota and
NLS-Maxwell-Bloch equations, a novel rational solution of the MNLS
equation is given as follows
\begin{equation}\label{firstrw}
{q^{[1]}}=-\exp[i(-x-tb-t+ta)]\dfrac{r_1r_2}{{r_1^*}^{2}},
\end{equation}\vspace{-0.8cm}
\begin{eqnarray*}
&r_1&\mbox{\hspace{-0.25cm}}=X+1+2ia[(3a-2)t-x],\\
&r_2&\mbox{\hspace{-0.25cm}}=X-3+2i[(-6a+3a^2+4b)t-ax],\\
&X&\mbox{\hspace{-0.25cm}}= 2(a-b)[(-2b+3a^2
\mbox{\hspace{-0.01cm}}-6a+4)t^2\mbox{\hspace{-0.01cm}}+4(1-a)xt+x^2].
\end{eqnarray*}
We omit the tedious calculation of this solution. The validity of this
solution has been confirmed by  symbolic computation. By letting $x \rightarrow {\infty}$ and $ t \rightarrow {\infty}$ it is easy to show that
 $|q^{[1]}|^{2} \rightarrow 1$, and $|q^{[1]}|^{2}(0,0)=9$. So, $q^{[1]}$ denotes a rational
solution on a non-zero background with a unit height. This  is a new
and wide class of solutions for the MNLS equation because it can
encompass five kinds of solution.

\noindent{\bf Transition between five states.} We are now in the
position to explore the properties of $q^{[1]}$ with more details.
The trajectory of $|q^{[1]}|^2$ in the $(x,t)$\/-plane is defined by
the location of the ridge (or vale) of its profile. A good
approximation of the trajectory for $|q^{[1]}|^2$ is given through a
simple equation $X+1=0$ in general. By a straightforward but tedious
calculation of the stationary point of $|q^{[1]}|^2$ in the
($x,t$)-plane we find, that, $q^{[1]}$ describes five solutions
associated with  five regions in the upper-right quadrant on the
($a,b$)-plane (see Figure 1), as follows \vspace{0.2cm}

\noindent
{\bf I} ( $b>a$). There is only one saddle point of $|q^{[1]}|^2$ at $(0,0)$, and there are
two simultaneous trajectories $X_1$ and $X_2$ on the ($x,t$)-plane. Note that the
 trajectories are not two straight lines as usual in the case of double solitons.
 If the height of soliton $|q^{[1]}|^2$ is increasing as it evolves along $X_1$, then it
  will decrease as it evolves on $X_2$. The asymptotic height of $|q^{[1]}|^2$ is
  $H_1=H_1(a,b)$ as $t=-\infty$\/, and $H_2=H_2(a,b)$ as $t=+\infty$ on $X_1$;
  $|q^{[1]}|^2$ approaches to $H_2$ as $t=-\infty$ and  to $H_2$ as $t=+\infty$ on
  $X_2$\/. Thus, when $(a,b) \in $ region
I, $q^{[1]}$ is called a paired bright-bright soliton because
$H_2>H_1>1$ and the appearance of two peaks (i.e. an increasing peak
and a decreasing one). Obviously, there exists energy exchange
between the two bright peaks propagating along $X_1$ and $X_2$ in
accordance with the energy conservation. In particular, the distance
between the two peaks is proportional to
$\sqrt{\delta_1t^2+\delta_2}$ in contrast to a linear function of
$t$ for the known two peaks in the case of a double soliton. Here
$\delta_1$ and $\delta_2$ are two real functions of $a$ and $b$.
\vspace{0.2cm}

\noindent {\bf II} ($b=a$). $|q^{[1]}|^2$ takes its maximum value 9
when ($x,t$) is on the line $x=(-2+3a)t$\/. This line is also the
trajectory of $|q^{[1]}|^2$\/. This is a single soliton solution.
\vspace{-0.05cm}

\noindent {\bf III} ($a > b > \max[0,a-\frac{3}{8}a^2]$). There is
only one extremum of $|q^{[1]}|^2$ at ($0,0$) on the ($x,t$)-plane.
Most features of the obtained solution are similar to those of
region I except $H_2>1>H_1>0$. Thus, in this region, $q^{[1]}$ is
called a paired bright-grey soliton. Note that $a-\frac{3}{8}a^2 <
0$ if $a>\frac{8}{3}$. \vspace{0.2cm}

\noindent {\bf IV} ($b=a-\frac{3}{8}a^2$\/, $a\in [0,\frac{8}{3}]$).
There is only one extremum of $|q^{[1]}|^2$ at ($0,0$) in the
($x,t$)-plane. The solution resembles that of region III except that
$H_2>1>H_1\geq0$. Thus, in region IV of the ($a,b$)-plane, $q^{[1]}$
is called a paired bright-dark soliton. \vspace{0.2cm}

\noindent {\bf V} ($0 < b < a-\frac{3}{8}a^2$). For $|q^{[1]}|^2$,
there is only one maximum at (0,0), where $|q^{[1]}|^2 = 9$, and two
minima, where $|q^{[1]}|^2 = 0$. The two minima are located with the
coordinates given by
$$
\begin{array}{ll} \displaystyle
x=\mp\dfrac{-6a+3a^2+4b}{a-b}\sqrt{\dfrac{3}{32((a-\frac{3}{8}a^2)-b)}},
   \vspace*{2mm}\\
\displaystyle t=\pm\dfrac{a}{a-b}\sqrt{\dfrac{3}{32((a-\frac{3}{8}a^2)-b)}}
\end{array}
$$
at {the points in the $(x,t)$\/-plane. When $(a,b)$ is} region V,
$|q^{[1]}|^2$ is localised both in the $x$ and $t$ direction, and
thus $q^{[1]}$ is a rogue wave solution of the MNLS
equation.\vspace{0.2cm}

The main difference between the grey and dark soliton is that the
minimum of the solution may or may not reach zero \cite{ablowitz1}
for the grey one.

For a physical system modelled by the MNLS equation, each solution
presented above gives a particular phase state. So, it is rather
interesting to observe in Figure~1 that there exists the state
transition induced by tuning the self-steepening and SPM, which can
be realised by adjusting the values of $a$ and $b$\/. For example,
by setting $a=\frac12$ and varying $b$\/, the system will evolve
consecutively through a paired bright-bright, a single, a paired
bright-grey, a paired bright-dark soliton states, and the rogue wave
state as $b$ decreases from a value larger than $\frac{1}{2}$ to one
that is smaller than $\frac{13}{32}$. Moreover, for a given
$b>b_c(=\frac{2}{3})$, the system passes through the first three
(i.e.\ I-III) states as $a$ increases from $zero$ to a sufficiently
large value; for $b=b_c$, the system now passes through the first
four (i.e. I-IV) states, and it passes through the paired
bright-grey soliton state twice; finally, for $b<b_c$, the system
passes through all five states and will be in paired bright-grey and
paired bright-dark soliton states twice. This newly discovered
unique state transition phenomenon is not described by the rational
solutions of the NLS and DNLS equations because  they do not
describe two different nonlinear effects  simultaneously. Thus, we
have now solved the long-standing problem mentioned in the
Introduction.

\noindent{\bf Particular cases.} In what follows we use two methods
to visualise function $|q^{[1]}|^2$\/. The first method consists of
plotting the profile of $|q^{[1]}|^2$ which is the graph of this
function of two variables in three dimensions. The second method is
similar to drawing the level lines of this function, however, not
using the lines but various colours instead. The figure obtained
this way is called the density plot of the function $|q^{[1]}|^2$\/.
In this plot each colour corresponds to a definite value of
$|q^{[1]}|^2$\/.

To illustrate the general results we investigate the evolution of
the rational solution for a fixed $a=\frac12$ and varying $b$\/. In
Figure~2 the profile (left panel) and density plot (right panel) of
$|q^{[1]}|^2$ are shown for a paired bright-bright soliton with
$b=1$. In Figure 3 the energy exchange (left panel) for two peaks
along the two trajectories (right panel) is plotted. The limit
heights in Figures 2 and 3 are $H_2=21+4\sqrt{5}\approx 29.9$ and
$H_1=21-4\sqrt{5}\approx 12.1$. Note that the height of the
background in Figure 2 is equal to 1. On the right panel of Figure
3, the explicit equations of trajectories $X_1$ (red line) and $X_2$
(green line) are $-t+\frac{1}{2}\sqrt{5t^2+4}=x$ and
$-t-\frac{1}{2}\sqrt{5t^2+4}=x$, respectively. The distance between
the two peaks at a given time is $\sqrt{5t^2+4}$ unlike the case of
the two peaks in a double soliton solution. It is  straightforward
to see in Figure 3 (left panel) that the energy is transmitted
gradually from a bright soliton (green line) moving along $X_2$ to
the other bright soliton (red line)  moving along $X_1$\/. The
reflective symmetry of Figure 3 (left panel) refers to the energy
conservation . By comparing Figure 2 (right panel) with Figure 3
(right panel), we can now see that the trajectory gives a very good
approximation of the ridge location in the profile of
$|q^{[1]}|^2$\/. Setting $a = b = \frac12$,  so that ($a, b$) is in
region II, we obtain $|q^{[1]}|^2=\dfrac{36+(t+2x)^2} {4+(t+2x)^2}$,
which is a single soliton with the height 9 and the exact trajectory
$t + 2x = 0$. We do not show the profile in this case because it is
a standard soliton.

In region III, by setting $b = \frac9{20}$, we obtain a paired
bright-grey soliton. Its profile is  plotted in Figure 4 (left
panel). The limit height of the bright soliton is
$H_2=\frac{8-\sqrt{15}}{20-5\sqrt{15}}\approx 6.5$, and the limit
height of the grey soliton is $H_1
=\frac{8+\sqrt{15}}{20+5\sqrt{15}} \approx 0.3$. The $X_1$ (red
line) and $X_2$ (green line) in Figure 4 (right panel) give good
approximation of the trajectories for $|t|> 8$.

Let us set $a = \frac12$ and $b=a-\frac38 a^2 \approx 0.41$,  the
point $(a,b)$ is now in region IV, and the rational solution
describes a bright-dark soliton. The profile of this solutions is
shown in Figure~5 (left panel). The limit heights are $H_2=4$ for
the bright soliton and $H_1=0$ for the dark one. The trajectories
$X_1$ and $X_2$ are shown on the left panel of Figure~5 by the red
and green lines, respectively. The analysis of the two solutions
corresponding to the bright-grey and bright-dark solitons shows
that: (i) There is energy exchange between the bright and grey
(dark) solitons; (ii) the bright soliton is transformed into a grey
(dark) one because its energy is lost during the interaction, while
the grey (dark) soliton is transformed into a bright one for $t \gg
1$.

Finally, by setting $a = \frac12$ and $b = \frac13$, we put the
 point $(a,b)$ in region V. In this case the rational solution describes
 the first-order rogue wave, which is plotted in Figure 6. 

\noindent{\bf Conclusions.} We have presented a range of new types
of rational solution of the MNLS equation that describe the
propagation of e.g. Alfv\'en waves in magnetised plasmas and the
femtosecond light pulses in optical fibres. The obtained solutions
have two free parameters, $a$  and $b$, representing the
contributions of  self-steepening and self phase-modulation, and,
depending on the values of these parameters, these solutions
describes five types of novel solitons corresponding to five states
of an associated physical system. These solutions are:  A paired
bright-bright, single, paired bright-grey, and a paired bright-dark
soliton, and a rogue wave. We have found that the state transition
among these five states is induced by  tuning the effects of
self-steepening and SPM. We urge that this novel phenomenon may be
observed in laboratory or in magnetised plasma in nature in order to
demonstrate an intricate balance between the effects of
self-steepening and SPM in an associated physical system.
Furthermore, because of the recent discovery of Alfv\'en  waves (see
\cite{alfven,robertus1} and references therein) in the magnetised
solar atmosphere, it is now paramount interest to find these novel
states and the state transfer in space plasmas, and, to establish
their connection with the long-standing coronal heating problem
\cite{clare,robertus2,robertus3,morton}.

Finally, we discuss briefly the novelty of the solution describing
the paired bright-bright soliton. Three of its characteristics, i.e.
the existence of two peaks, decreasing or increasing amplitude, and
the curved trajectories, are essentially different from the similar
characteristics of the recently found solition solutions that
include an explode-decay soliton \cite{nakamura}, a two-peak soliton
\cite{satsuma,li}, a W-shape soliton \cite{zhou2}, a dark-in-bright
soliton \cite{Kevrekidis,prosezian}, a rogue wave \cite{peregrine}
and a two-peak rogue wave \cite{he2012,akhmedievpre2012}, in
addition to the well-known classical soliton, breather, and kink. In
particular, the paired bright-bright soliton is not a travelling
wave. The non-autonomous solitons \cite{serkin1,jvv1,jvv2,heli2011}
have the properties similar to those of the paired bright-bright
soliton but only  when they are solutions to the variable
coefficient soliton equations. This discussion can also be applied
to the paired bright-grey and bright-dark soliton solutions.\\
\noindent{\bf Acknowledgements} This work is supported by the NSF of
China Grant No.10971109 and 11271210, and K.C. Wong Magna Fund in
Ningbo University. J. He is also supported by the Natural Science
Foundation of Ningbo, Grant No. 2011A610179. R.E. acknowledges M.
K\'eray for patient encouragement and is also grateful to NSF,
Hungary (OTKA, Ref. No. K83133) for the support received. J. He
thank sincerely Prof. A.S. Fokas for arranging the visit to
Cambridge University in 2012-2013 and for many useful discussions .
 \mbox{}
\begin{figure}[htb]
\includegraphics*[height=3.5cm]{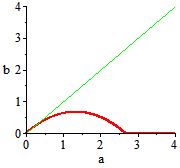}
\caption{(Colour online) Five regions in the upper-right quadrant on
the ($a,b$)-plane. In each region, $q^{[1]}$ gives a new kind of
solution for the MNLS equation. The straight green line is $a = b$,
the red curve is defined by $b = a - \frac38 a^2$ for $a \leq
\frac83$, $b = 0$ for $a \geq \frac83$\/.}
\end{figure}
\begin{figure}[htb]
\includegraphics*[height=3.6cm]{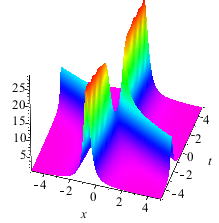}
\includegraphics*[height=3.6cm,width=3.6cm]{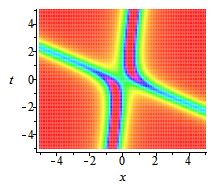}
\mbox{}\vspace{-.3cm} \caption{(Colour online) The profile (left)
and density plot (right) for a paired bright-bright soliton.}
\end{figure}
\begin{figure}[htb]
\includegraphics*[height=3.6cm]{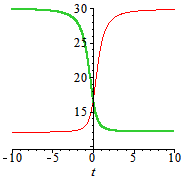}
\includegraphics*[height=3.6cm]{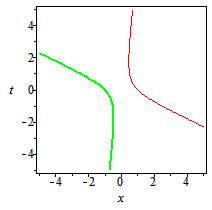}
\mbox{}\vspace{-.3cm} \caption{(Colour online) Left: The energy
transmission from a bright soliton (green line) to another one (red
line). Right: The trajectories of the solitons, $X_2$ (green line)
and $X_1$ (red line).}
\mbox{}\vspace{-0.5cm}
\end{figure}
\begin{figure}[htb]
\includegraphics*[height=4.5cm]{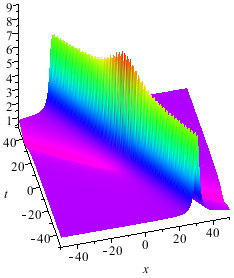}
\includegraphics*[height=3.1cm]{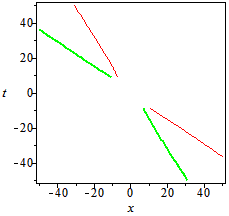}
\mbox{}\vspace{-.3cm} \caption{(Colour online) The paired
bright-grey soliton (left) and its trajectories (right).}
\mbox{}\vspace{-0.5cm}
\end{figure}
\begin{figure}[htb]
\includegraphics*[height=3.6cm]{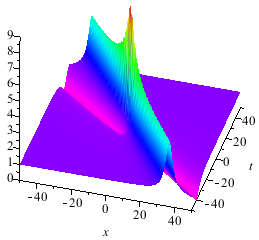}
\includegraphics*[height=3.2cm]{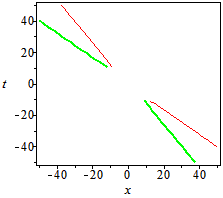}
\mbox{}\vspace{-.3cm} \caption{(Colour online) The paired
bright-black soliton (left) and its trajectories (right).}
\mbox{}\vspace{-0.5cm}
\end{figure}
\begin{figure}[htb]
\includegraphics*[height=4.1cm]{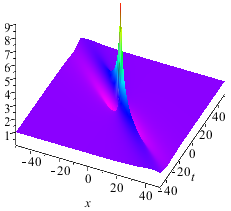}
\includegraphics*[height=3.3cm]{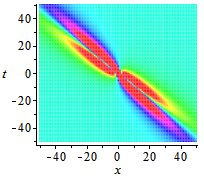}
\mbox{}\vspace{-.3cm} \caption{(Colour online) The first-order rogue
wave (left) and its density plot (right).}
\end{figure}

\vspace{-0.5cm}
\begin{thebibliography}{200}
\bibitem{KMio}
K. Mio, T. Ogino, K. Minami and S. Takeda, J. Phys. Soc. Jpn.
41(1976), 265-271.
\bibitem{medvedev}M. V. Medvedev, P. H. Diamond, V. I Shevchenko  and V. L. Galinsky
Phys. Rev. Lett. 78(1997), 4934-4937.
\bibitem{sulem} D. Laveder, T. Passot, P.L. Sulem, Phys. Lett.
A377(2013),1535-1441.
\bibitem{Tzoar} N. Tzoar and M Jain, Phys. Rev. A 23(1981),1266-1270.
\bibitem{Anderson}D. Anderson and M. Lisak, Phys.Rev. A 27(1983),1393-1398
\bibitem{Zabolotskii} A. A. Zabolotskii,Phys. Lett. A. 124(1987), 500-502.
\bibitem{MikiWadati1} M. Wadati, K. Konno and Y.H. Ichikawa,
 J. Phys. Soc. Jpn. 46(1979), 1965-1966.
\bibitem{AKundu}
A. Kundu, J. Math. Phys. 25(1984),3433-3438.
\bibitem{Seenuvasakumaran} K. Porsezian,
P. Seenuvasakumaran and K. Saravanan,Chaos, Solitons and Fractals.
11(2000), 2223-2231.
\bibitem{QingDing}
Q. Ding and Z.N. Zhu,Phys. Lett. A. 295(2002), 192-197.
\bibitem{Kundu}A.Kundu,Symmetry, Integrability and Geometry:
Methods and Applications. 2(2006), 078(12pp).
\bibitem{TKawata}T.Kawata, J.I.Sakai and N.Kobayashi,
 J.Phys. Soc. Jpn. 48(1980), 1371-1379.
\bibitem{Chowdhury}
A. R. Chowdhury, S. Paul  and S. Sen, Phys. Rev. D. 32(1985),
3233-3237.
\bibitem{mihalache} D.Mihalache, N.Truta, N.C.Panoiu and
D.M..Baboiu, Phys. Rev. A. 47(1993), 3190-3194.
\bibitem{Doktorov2}E. V. Doktorov,  Eur. Phys. J. B. 29(2002), 227-231.
\bibitem{Rangwala}
A.A.Rangwala and J.A. Rao, J. Math. Phys. 31(1990), 1126-1132.
\bibitem{OCWrigh}
O.C. Wrigh,  Chaos, Solitons and Fractals. 20(2004), 735-749.
\bibitem{TiechengXia}
T.C. Xia, X.H.Chen and D.Y.Chen, Chaos, Solitons and
Fractals. 26(2005), 889-896.
\bibitem{HaiQiangZhang}H.Q.Zhang, B.G.Zhai and X.L. Wang,
Phys. Scr. 85(2012),
015007(8pp).

\bibitem{Shanliang} S.L. Liu and W.Z.Wang, Phys. Rev. E. 48(1993), 3054-3059.
\bibitem{MinLi}M.Li, B. Tian, W.J.Liu, H.Q.Zhang and P.Wang, Phys. Rev.
E. 81(2010), 046606(8pp).
\bibitem{serkindecay} E.A.Golovchenko,E.M.Dianov,A.M.Prokhorov and V.N.Serkin,
JETP Lett.42(1985), 87-91.
\bibitem{perturbation}
X. J. Chen and J. K. Yan,Phys. Rev. E.65(2002), 066608 (12pp).
\bibitem{Shchesnovich}
V.S. Shchesnovich and E.V. Doktorov, Physica. D. 129( 1999),
115-129.
\bibitem{VMLashkin}
V. M. Lashkin,  Phys. Rev. E. 69(2004),  016611(11pp).
\bibitem{he2}J.S.He, H.R. Zhang, L.H. Wang, K. Porsezian and  A.S.Fokas,
A generating mechanism for higher order rogue
waves(arXiv:1209.3742v4).
\bibitem{xuhe}S.W. Xu, J.S.He and L.H.Wang, J. Phys. A: Math. Theor.
44(2011), 305203 (22pp).
\bibitem{xxuhe}S.W. Xu and J.S.He, J. Math. Phys. 53(2012), 063507 (17pp)
\bibitem{taohe}Y.S.Tao and J.S.He, Phys.Rev.E 85(2012), 026601(7pp).
\bibitem{hexup}J.S.He, S.W.Xu and K.Porsezian, Phys. Rev.E 86 (2012),066603(17pp).
\bibitem{ablowitz1}M.J. Ablowitz,2011, Nonlinear Dispersive Waves:
 Asymptotic Analysis and Solitons ( Cambridge University Press, Cambridge )p.153.
\bibitem{alfven}H. Alfv\'en, Nature 150(1942), 405-406.
\bibitem{robertus1} D. B. Jess, M. Mathioudakis, R. Erd\'elyi,
 P. J. Crockett, F. P. Keenan and D. J. Christian, Science 323(2009),1582-1585.
\bibitem{clare} C. E. Parnell and I. De Moortel, Phil. Trans. R. Soc. A 370(2012),3217-3240
\bibitem{robertus2}M. Mathioudakis, D. B. Jess and R. Erd\'elyi,
Alfv\'en Waves in the Solar Atmosphere: From Theory to Observations
(arXiv:1210.3625)29pp.
\bibitem{robertus3}S. Wedemeyer-B\"ohm, E. Scullion, O. Steiner,
L. R. van der Voort, J. de la Cruz Rodriguez, V. Fedun and  R.
Erd\'elyi, Nature 528(2012),505-508.
\bibitem{morton}R. Morton, G. Verth, D.B. Jess,  D. Kuridze, M.S. Ruderman, M. Mathioudakis,
and R. Erd\'elyi, Nature Comm, 3(2012), i.d.1315.
\bibitem{nakamura} A. Nakamura, J. Phys. Soc. Jpn. 50(1981), 2469-2470;ibid., 51(1982),19-20.
\bibitem{satsuma}N.Sasa and J.Satsuma, J. Phys. Soc. Jpn. 60(1991), 409-417.
\bibitem{li}Y.S.Li and W.T.Han, Chin.Ann.Math.22B(2001),171-176.
\bibitem{zhou2} Z.H.Li, L.Li, H.P.Tian and G.S.Zhou, Phys. Rev. Lett.84 (2000),
4096-4099.
\bibitem{Kevrekidis}
P. G. Kevrekidis, D. J. Frantzeskakis, B. A. Malomed,A. R. Bishop and I. G. Kevrekidis,
New.J.Phys.5(2003),64.
\bibitem{prosezian}A. Choudhuri and K. Porsezian, Optics Communications 285(2012), 364-367.
\bibitem{peregrine}D. H. Peregrine, J. Aust. Math. Soc. Ser. B: Appl. Math. 25(1983),
16-43.
\bibitem{he2012}J.S.He, S.W.Xu and  K.Porsezian,
J. Phys. Soc. Jpn. 81(2012), 033002.
\bibitem{akhmedievpre2012}U.Bandelow and N.Akhmediev, Phys. Rev. E 86 (2012), 026606.
\bibitem{serkin1}V.N. Serkin, A. Hasegawa and T.L. Belyaeva, Phys. Rev. Lett.98 (2007), 074102.
\bibitem{jvv1} J. Belmonte-Beitia, V. M. Perez-Garcia,V. Vekslerchik and
P. J. Torres, Phys. Rev. Lett. 98 (2007),064102.
\bibitem{jvv2} J. Belmonte-Beitia, V. M. Perez-Garcia, V. Vekslerchik and V. V. Konotop,
 Phys. Rev. Lett.100(2008),164102.
\bibitem{heli2011} J.S. He and  Y.S. Li, Stud. in Appl. Math.126(2011),1-15.
\end{thebibliography}
\end{document}